\documentclass[reprint,superscriptaddress]{revtex4-1}
\usepackage[utf8]{inputenc}

\usepackage{amsmath}
\usepackage{bm}
\usepackage{graphicx}
\usepackage{sidecap}
\usepackage{xcolor}
\usepackage{hyperref}
\usepackage{enumitem}

\DeclareMathOperator{\sech}{sech}
\newcommand{\p}{\partial}
\newcommand{\half}{\frac{1}{2}}

\newcommand{\e}{\bm{\hat{e}}}

\newcommand{\emn}{\epsilon_{\mu\nu}}

\newcommand{\Spin}{\bm{S}}
\newcommand{\magn}{\bm{m}}
\newcommand{\nagn}{\bm{n}}
\newcommand{\Heff}{\bm{F}}

\newcommand{\DM}{D}
\newcommand{\dm}{\lambda}

\newcommand{\Anisotropy}{k}

\newcommand{\Energy}{E}
\newcommand{\Potential}{V}
\newcommand{\Eex}{E_{\rm ex}}
\newcommand{\Edm}{E_{\rm DM}}
\newcommand{\Edmm}{\tilde{E}_{\rm DM}}
\newcommand{\Ean}{E_{\rm an}}
\newcommand{\damp}{\alpha}

\begin{document}

\title{Breathing skyrmions in chiral antiferromagnets}
\author{S. Komineas}
\affiliation{Department of Mathematics and Applied Mathematics, University of Crete, 70013 Heraklion, Crete, Greece}
\affiliation{Institute of Applied and Computational Mathematics, FORTH, Heraklion, Crete, Greece}
\author{P. E. Roy}
\affiliation{Hitachi Cambridge Laboratory, Hitachi Europe Limited, Cambridge CB3 0HE, United Kingdom}
\date{\today}

\begin{abstract}
Breathing oscillations of skyrmions in chiral antiferromagnets can be excited by a temporal modification of the Dzyaloshinskii-Moriya interaction or magnetocrystalline anisotropy strength.
We employ an adiabatic approximation and derive a formula for the potential that directly implies breathing oscillations.
We study the nonlinear regime and the features of larger amplitude oscillations.
We show that there is a maximum amplitude supported by the potential.
As a consequence, we predict theoretically and observe numerically skyrmion annihilation events due to excitation of large amplitude breathing oscillations.
The process is efficient when the skyrmion is mildly excited so that its radius initially grows, while the annihilation event is eventually induced by the internal breathing dynamics.
We reveal the counter-intuitive property that the skyrmion possesses a nonzero kinetic energy at the moment of its annihilation.
Finally, the frequency of small amplitude breathing oscillations is determined.
\end{abstract}

\maketitle

\section{Introduction}
\label{sec:intro}

Magnetic skyrmions are topological textures with a swirling configuration of the magnetization stabilized by the Dzyaloshinskii-Moriya (DM) interaction \cite{BogdanovYablonskii_Eksp_Theor_Fiz,BogdanovHubert_JMMM1994}.
They are commonly observed in chiral ferromagnetic films, typically extending a few or tens of nanometers laterally, and their nontrivial topology makes them robust against perturbations.
Skyrmions exhibit particle-like dynamics \cite{2015a_PRB_KomineasPapanicolaou} which, together with their small size, lead to properties such as a low driving threshold current \cite{JIANG20171}.
These qualities have given rise to a range of proposed applications such as their use as the constituent information carriers in racetrack memory/logic devices \cite{Fert_Nat_2017,TomaselloSciRep_2014,Koshibae_2015,Zhang_SciRep_2015,LuoNanoLett_2018}, memristor elements in artificial synapses for neuromorphic computing architectures \cite{Song_NatElec_2020}, in spintronics-based transistor device concepts \cite{Zhang2_SciRep_2015}, and as spin-wave scatterers in magnonic computing and logic devices \cite{MoonPhysRevApplied2016}.

Most of the existing work is focused on ferromagnetic skyrmions. 
On the other hand, skyrmions are expected to exist also in antiferromagnets (AFM) for essentially the same reasons as in ferromagnets.
Antiferromagnetic materials have advantages such as high operational frequencies in the THz range and robustness against external magnetic field perturbations.
Furthermore, unlike in ferromagents and despite their topological nature, skyrmions in AFMs do not present a Hall angle in their dynamical behavior.
These properties have prompted proposals to replace ferromagnetic skyrmions by their antiferromagnetic counterparts \cite{BarkerPhysRevLett2016,JinAPL_2016,XiaJPhysD_2017,XiaofengAPL_2018,PhysRevBLiang2019,JinPhysRevB2020,Zhang2_SciRep_2015}.
Recent experimental observations of stable skyrmions in AFMs \cite{GaoNat2020,JaniNat2021} give further motivation for the study of AFM skyrmions for future device implementations.

The presence of the chiral DM interaction is crucial for the dynamics of solitons, in addition to its role in their stabilization.
A skyrmion breathing mode arises in chiral ferromagnets \cite{2014_PRB_SchuetteGarst} due to the breaking of the conservation of the total magnetization perpendicular to the film.
A similar mode exists also for antiferromagnetic skyrmions.
In the latter case, independent oscillations of the skyrmion radius and of its chirality are possible \cite{2019_PRB_KravchukGomonaySheka}.

We are motivated by the emerging kinetic energy in the AFM continuum \cite{BaryakhtarChetkin1994,KomineasPapanicolaou_NL1998} and we propose an effective potential for the skyrmion oscillation dynamics.
This leads to a systematic approach in order to understand details of these dynamical modes, such as the frequency of oscillations, by a combination of analytical and numerical methods.
The form of the effective potential leads to the observation that an {\it expansion} of the skyrmion can eventually lead to its annihilation by a subsequent collapse due to internal breathing dynamics.
This counter-intuitive method shows that a large external force is {\it not} necessary in order to annihilate the topological texture; instead skyrmion annihilation is obtained by small controlled changes of the DM or anisotropy parameters that can be induced, e.g., by a voltage pulse \cite{2021_JMMM_Schott}, in combination with internal dynamics.

It would be an ideal situation to have methods for skyrmion creation and annihilation by controlled small perturbations while at the same time the skyrmion remains robust to all usual perturbations.
If this is achieved the potential for skyrmions as functional objects would be significantly enhanced.
In this context, our method shows the way for controlled skyrmion annihilation by small perturbations.
In addition, the successful analytical arguments presented may be useful in studies for a method towards controlled individual skyrmion generation using only mild forces.

Sec.~\ref{sec:breathing} discusses the energy in an antiferromagnet in the discrete and in the continuum model.
Sec.~\ref{sec:annihilation} shows how skyrmion annihilation via breathing dynamics can be achieved.
In Sec.~\ref{sec:nonlinear}, we derive the effective potential for breathing oscillations and give a theoretical description of the annihilation dynamics.
In Sec.~\ref{sec:smallOscillations}, we derive the frequency of small amplitude breathing oscillation in the case of skyrmions with a large and a small radius.
Sec.~\ref{sec:helicity} discusses helicity oscillations for the skyrmion.
Sec.~\ref{sec:conclusions} contains our concluding remarks.

\section{Skyrmion energy}
\label{sec:breathing}

We consider a square lattice of spins in a material with the usual exchange, perpendicular anisotropy and Dzyaloshinskii-Moriya interaction.
The magnetic energy of the lattice is
\begin{equation} \label{eq:energyDiscrete}
\begin{split}
    \Energy^d = & \sum_{i,j} J_n\, \Spin_{i,j}\cdot (\Spin_{i+1,j} + \Spin_{i,j+1}) 
    + \frac{\Anisotropy}{2}\, [1 - (\Spin_{i,j})_3^2] \\
    & + \DM\, [ \e_2\cdot (\Spin_{i,j}\times\Spin_{i+1,j}) - \e_1\cdot (\Spin_{i,j}\times\Spin_{i,j+1})]
\end{split}
\end{equation}
where the spin variables are assumed normalized $|\Spin_{i,j}|=1$.
We will typically use, in numerical simulations, the parameter values
\begin{equation} \label{eq:parameterValues}
    J_{n} = 2\times\text{10}^{\text{-21}}\,{\rm Joule},\quad
    \DM = 0.047 J_{n},\quad
    \Anisotropy = 0.01 J_{n}.
\end{equation}
The equation of motion for the spins is
\begin{align} \label{eq:Heisenberg}
& \frac{\p \Spin_{i,j}}{\p t} = -\gamma\Spin_{i,j}\times\Heff_{i,j} + \damp \Spin_{i,j}\times\frac{\p \Spin_{i,j}}{\p t},  \\
& \Heff_{i,j} = - \frac{1}{\mu_0\mu_s}\frac{\p \Energy^d}{\p\Spin_{i,j}} \notag
\end{align}
where $\Heff$ is the effective field, $\gamma=g_e \mu_B \mu_0/\hbar = 2.211\times10^5\, {\rm m}\,{\rm A}^{-1} {\rm s}^{-1}$ is the gyromagnetic ratio, and $\damp$ is the damping parameter.
We choose the saturation magnetization $\mu_s=4\mu_B$, where $\mu_B$ is the Bohr magneton.
The material parameters have been chosen to resemble what is expected for a range of antiferromagnetic oxides \cite{2011_Archer,2017_Gopal}.

A continuum model is obtained if we consider a lattice of spin tetramers where the normalized N\'eel vector $\nagn_{\alpha,\beta}$ is defined at tetramer sites $(\alpha, \beta)$ \cite{KomineasPapanicolaou_NL1998}.
The distance between tetramer sites is defined to be $2\epsilon$ where $\epsilon\equiv \sqrt{\Anisotropy/J}$ is a small parameter.
In the limit $\epsilon\to 0$, a continuous N\'eel vector field $\nagn=\nagn(x,y,\tau)$ is obtained, with $\nagn^2=1$, where $x,y$ and $\tau$ are scaled space and time variables \cite{BaryakhtarIvanov_SJLTP1979,BaryakhtarChetkin1994,KomineasPapanicolaou_NL1998,GomonayLoktev_PRB2010}.
The energy in the continuum is (see also Refs.~\cite{BogdanovYablonskii_afm_JETP1989,BogdanovShestakov_PSS1998})
\begin{equation} \label{eq:energyContinuum}
    \Energy = T + \Potential_\dm
\end{equation}
where the kinetic energy is
\begin{equation} \label{eq:kinetic}
    T = \half \int \dot{\nagn}^2\,d^2x 
\end{equation}
and the dot denotes differentiation with respect to the scaled time $\tau$, and the potential energy
\begin{equation} \label{eq:potential}
\Potential_\dm = \Eex + \Edm + \Ean
\end{equation}
includes exchange, DM, and anisotropy contributions,
\begin{equation} \label{eq:potentialComponents}
\begin{split}
& \Eex = \half \int (\p_\mu\nagn)\cdot(\p_\mu\nagn)\, d^2x \\
& \Edm = \dm \int \emn \bm{\hat{e}}_\mu\cdot(\p_\nu\nagn\times\nagn)\, d^2x \\
& \Ean =\half \int (1-n_3^2)\, d^2x
\end{split}
\end{equation}
with $\dm=\DM/\sqrt{\Anisotropy J_n}$ being a scaled DM parameter, and all energy components being in units of $J_n$.
Symbols $\p_\mu, \p_\nu$, with $\mu,\nu=1,2$, denote differentiation with respect to $\left(x, y\right)$, respectively, $\emn$ is the antisymmetric tensor, and the summation convention for repeated indices is adopted.
Note that $x$ is a scaled coordinate and actual distances (in physical units) are given by
\begin{equation} \label{eq:physicalLength}
    a x/\epsilon,\qquad a y/\epsilon
\end{equation}
where $a$ is the distance between neighboring spins.
The unit of time is
\begin{equation} \label{eq:physicalTime}
    \tau_0 = \frac{\mu_s}{g_e \mu_B} \frac{\hbar}{2\sqrt{2\Anisotropy J_n}} =  0.373\,{\rm ps}
\end{equation}
where the numerical value corresponds to the parameter values in Eq.~\eqref{eq:parameterValues} and $\mu_s = 4\mu_B$.

The potential energy $\Potential_\dm$ is identical in form to the energy for a ferromagnet with corresponding interactions.
It is thus known that the ground state is uniform (N\'eel) for $\dm < 2/\pi$ while it is the spiral for $\dm > 2/\pi$ \cite{BogdanovHubert_JMMM1994}.
Isolated skyrmions are localised excitations on a uniform background.
On the other hand, the presence of the kinetic term in Eq.~\eqref{eq:energyContinuum} opens possibilities that are not there in ferromagnets. 
We will study oscillations of the skyrmion.

We consider an axially-symmetric skyrmion configuration for the N\'eel vector, written conveniently in terms of the spherical variables
\begin{equation}
    \Theta=\Theta(r,t),\quad \Phi = \phi + \chi(t)
\end{equation}
where $(r,\phi)$ are polar coordinates and $\chi$ is an angle that may be called the helicity.
We will assume that $\chi$ may depend on time only or it is a constant.
The kinetic energy of Eq.~\eqref{eq:kinetic} takes the form
\begin{equation} \label{eq:kinetic_spherical}
T = \frac{1}{2} \int \left( \dot{\Theta}^2 + \sin^2\Theta\, \dot{\chi}^2 \right) \, 2\pi r\,dr.
\end{equation}
The exchange and anisotropy energies depend on $\Theta$ only,
\begin{equation} \label{eq:energyExchange_skyrmion}
\begin{split}
    \Eex & = \frac{1}{2} \int \left[ (\Theta')^2 + \frac{\sin^2\Theta}{r^2} \right]\, 2\pi r\,dr, \\
    \Ean & = \frac{1}{2} \int \sin^2\Theta\,\, 2\pi r\,dr
\end{split}
\end{equation}
where the prime denotes differentiation with respect to the space variable $r$.
The DM term is written as
\begin{equation} \label{eq:energyDM_skyrmion}
\begin{split}
\Edm & = \dm \cos\chi\,\Edmm, \\
\Edmm & = \int \left( \Theta' + \frac{\cos\Theta \sin\Theta}{r} \right)\, 2\pi r\,dr.
\end{split}
\end{equation}

\begin{figure}[t]
\begin{center}
\includegraphics[width=8cm]{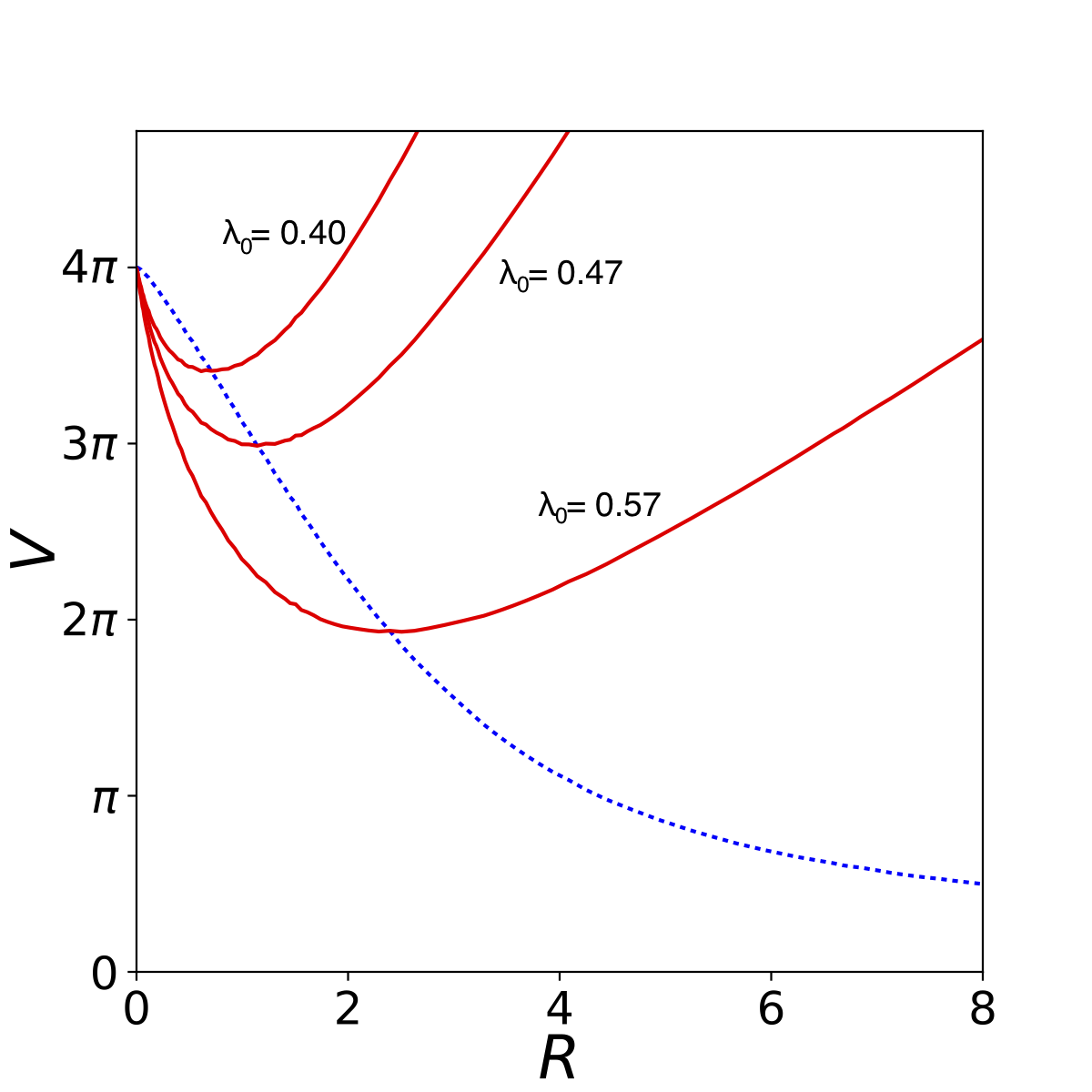}
\caption{
The blue dotted line shows the potential energy $\Potential_\dm$ for the static skyrmion versus its radius $R=R(\dm)$.
The red solid lines show the potential energy $V_{\dm_0}(\Theta_\dm)$ given in Eq.~\eqref{eq:potential_dm0} with $\dm_0=0.40, 0.47, 0.57$ and corresponding equilibrium radii $R_0 = 0.71, 1.14, 2.39$.
The radius $R$ is in scaled units and the actual length is given, as in Eq.~\eqref{eq:physicalLength}, by $a R/\epsilon$ (it is $\epsilon=0.1$ for parameter values in Eq.~\eqref{eq:parameterValues}).
}
\label{fig:breathingPotential}
\end{center}
\end{figure}

The static skyrmion profile that minimizes the potential energy $\Potential_\dm$ will be denoted by $\Theta_\dm(r)$ and it is obtained for $\chi=0$.
The skyrmion radius depends on the parameter $\dm$.
For small $\dm$ the skyrmion radius $R$ is small and for $\dm\to 0$ the radius $R\to 0$ and the energy takes the value $\Energy=4\pi$ \cite{KomineasMelcherVenakides_NL2020,BernandMuratovSimon_PRB2020,2020_ARMA_BernandMuratovSimon}.
The radius $R$ increases with $\dm$ and it diverges, $R\to\infty$, for $\dm\to 2/\pi$ \cite{RohartThiaville_PRB2013,KomineasMelcherVenakides_PhysD2021}.
Fig.~\ref{fig:breathingPotential} shows the numerically calculated energy from Eq.~\eqref{eq:potential} of a static skyrmion as a function of its radius $R$, defined to be at the point where the magnetization points in-plane, i.e., $\Theta=\pi/2$.

We conclude this section by introducing the local magnetization vector $\magn$ defined as the normalized mean value of the spins on tetramers \cite{KomineasPapanicolaou_NL1998,KomineasPapanicolaou_SciPost2020}.
It is an auxiliary field in the continuum theory, given by
\begin{equation}  \label{eq:magn0}
\magn = \frac{\epsilon}{2\sqrt{2}}\, \nagn\times\dot{\nagn}.
\end{equation}
A nonzero magnetization is obviously connected with dynamics.
The kinetic energy in Eq.~\eqref{eq:kinetic} is given in terms of the magnetization of Eq.~\eqref{eq:magn0} as
\begin{equation} \label{eq:kinetic_magnetization}
    T = \frac{4}{\epsilon^2} \int \magn^2\,d^2x.
\end{equation}
In the following, we will make use of its discretized form
\begin{equation} \label{eq:kinetic_magnetization_discrete}
    T = 16 \sum_{\alpha,\beta} \magn_{\alpha,\beta}^2
\end{equation}
that gives the kinetic energy as a sum over the lattice of tetramers.

\section{Skyrmion annihilation via breathing}
\label{sec:annihilation}


The fact that the energy of an infinitesimally small skyrmion is finite, $\Energy=4\pi$, as shown in Fig.~\ref{fig:breathingPotential}, suggests the possibility to annihilate the skyrmion by a finite force.
This idea will be combined with the dynamics of the breathing mode that is known to exist in chiral magnets \cite{2014_PRB_SchuetteGarst}, as will be explained in the following.

We assume a material with parameter value $\dm_0$ that gives a static skyrmion with radius $R_0$ and energy $\Potential_0$.
We further assume a method to expand this skyrmion and produce one with a larger radius $R > R_0$ and, naturally, a larger energy $\Potential > \Potential_0$.
We expect that the skyrmion energy, when this is out of equilibrium, can be chosen $\Potential > 4\pi$ if $R$ is large enough.
Using such a large radius skyrmion as an initial state, we anticipate that the breathing dynamics can reduce the skyrmion radius down to $R\to 0$ since the energy of the oscillator is larger than the potential energy of a $R\to 0$ skyrmion.


\begin{figure}[t]
    \centering
    \includegraphics[width=\columnwidth]{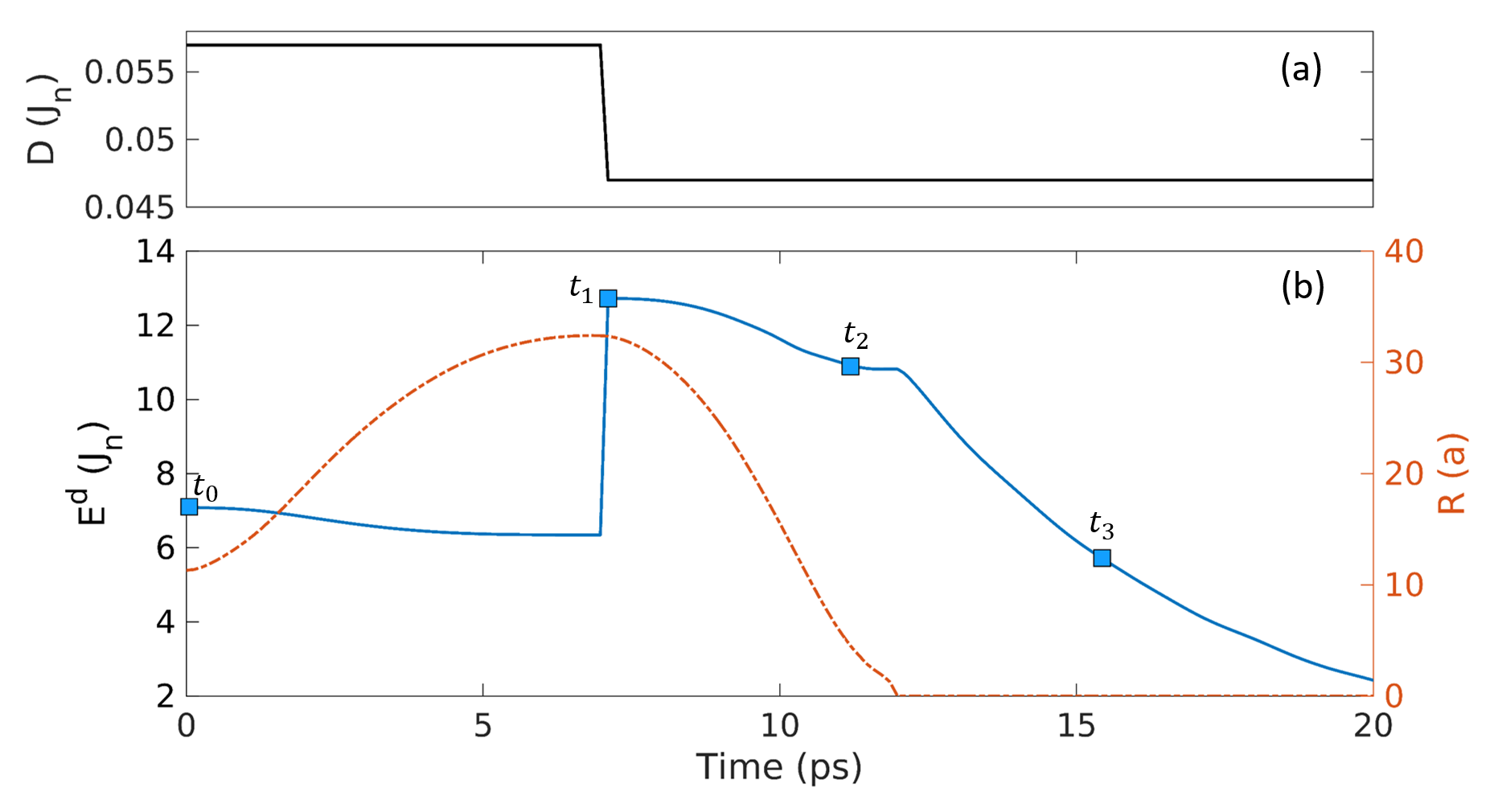}
    \includegraphics[width=\columnwidth]{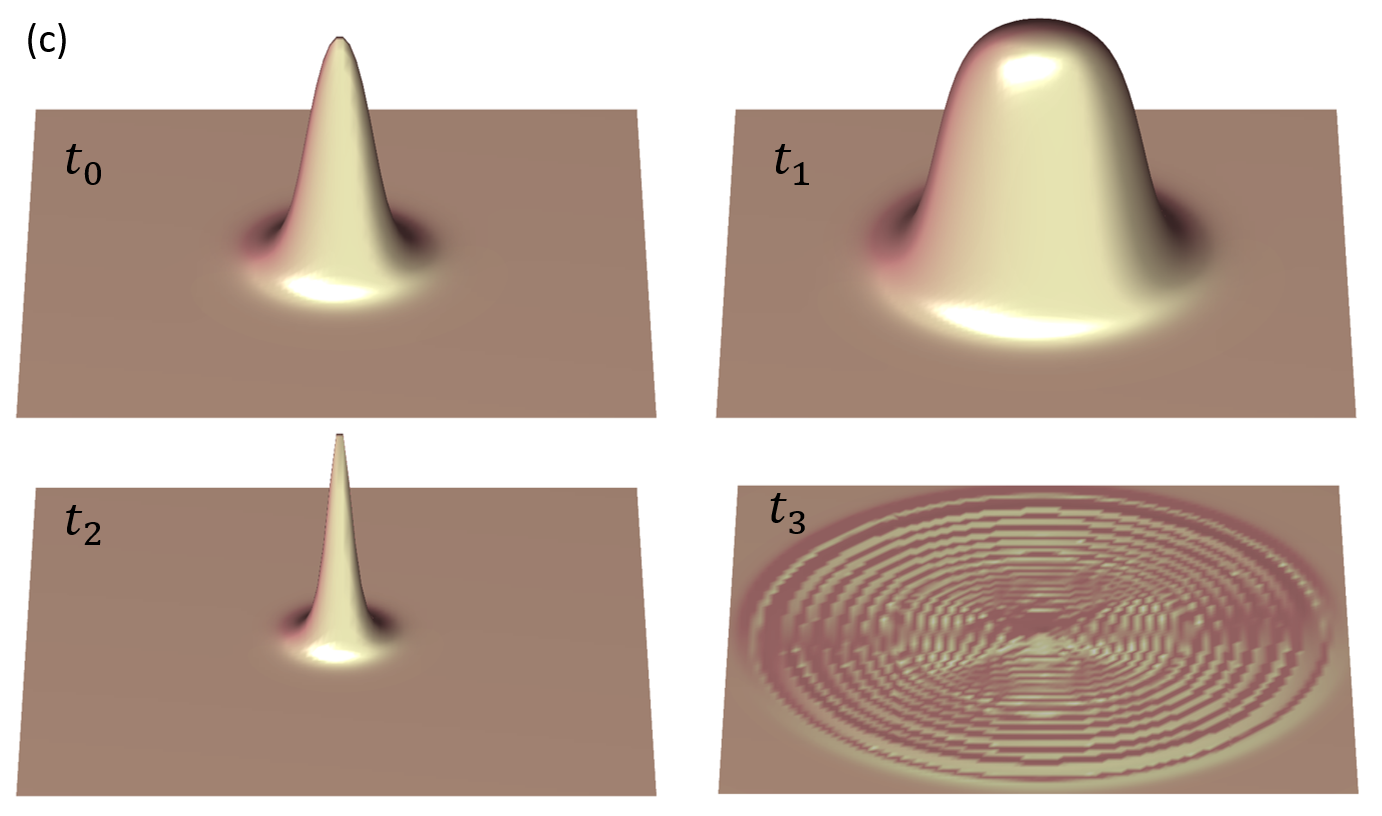}
    \caption{Skyrmion annihilation via breathing.
    (a) The parameter $\DM$ is shown vs time.
    A square pulse of 7 ps is modulating the DM parameter to the value $\DM=0.057\,J_n$ ($\dm=0.57$).
    After the pulse is switched off, it is $\DM=0.047\,J_n$ ($\dm=0.47$).
    (b) The total energy from Eq.~\eqref{eq:energyDiscrete} vs time is shown by a blue solid line.
    The skyrmion radius is shown by a red dashed line.
    It is given in units of lattice spacing $a$.
    (c) Spatial distribution of $n_3$ at time points $t_i$ (indicated on the energy graph).
    Snapshot $t_0$ shows the initial skyrmion (static profile for parameter values of Eq.~\eqref{eq:parameterValues}), snapshot $t_1$ shows the skyrmion when the pulse is switched off, snapshot $t_2$ shows the skyrmion before annihilation, snapshot $t_3$ shows the generated waves after the skyrmion has been annihilated.
    Only part of the simulation space is shown.}
    \label{fig:annihilationDynamics}
\end{figure}

We implement the above ideas in a numerical simulation of a spin lattice with $500\times500$ sites.
We consider a static skyrmion for the parameter values in Eq.~\eqref{eq:parameterValues}.
We propagate in time the dynamical equations \eqref{eq:Heisenberg} including damping with $\damp=0.0025$.
The system is subjected to a voltage pulse that initially modifies the DM parameter to the value $\DM=0.057\,J_n$ ($\dm=0.57$) as shown in Fig.~\ref{fig:annihilationDynamics}a.
The evolution of the skyrmion radius and discrete energy \eqref{eq:energyDiscrete} are shown in Fig.~\ref{fig:annihilationDynamics}b.
Four snapshot of the skyrmion profile ($n_3$ component) during the simulation are shown in Fig.~\ref{fig:annihilationDynamics}c.
At time $t_0$, the initial skyrmion is shown.
The skyrmion expands during the voltage pulse, and it is shown at the end of the pulse at time $t_1$.
Subsequently, breathing dynamics induces shrinking of the skyrmion, as shown at time $t_2$.
The skyrmion is eventually annihilated at time approximately $12\,{\rm ps}$.
After annihilation, low amplitude waves in the form of radiation are found to spread radially, away from the original skyrmion center, as shown at time $t_3$.
In Fig.~\ref{fig:annihilationDynamics}b, we see that the skyrmion radius is increasing during the voltage pulse.
The pulse is switched off at the time that the skyrmion has maximum radius.
Then, the radius is decreasing due to breathing dynamics until the skyrmion shrinks to a point and is annihilated.
The energy is decreasing due to damping.
There is a step-like increase of the energy when the voltage pulse is switched off, due to the drop of the DM parameter.
This brings the energy to a value greater than $4\pi$.
When annihilation of the skyrmion happens, the discrete energy is somewhat lower than $4\pi$ (which is the prediction of the continuum theory for minimum skyrmion energy).
Certainly, the prediction based on the continuum model is not expected to be quantitatively correct when the skyrmion is concentrated in a few lattice sites, just before annihilation.
The energy dissipation continues faster after the skyrmion annihilation.
We stop the simulation when radiation reaches the boundaries of the numerical mesh.

We proceed to a detailed study of the skyrmion annihilation dynamics.
At the time that the skyrmion is concentrated in a point, the kinetic energy should be equal to $T=\Potential - 4\pi > 0$.
This seems incompatible with the intuitive expectation that $\dot{R}\to 0$ at $R=0$.
The apparent contradiction can be resolved if we estimate the kinetic energy for a skyrmion of small radius where the profile is approximated by a Belavin-Polyakov (BP) configuration \cite{KomineasMelcherVenakides_NL2020,BernandMuratovSimon_PRB2020}.
A BP skyrmion with a time dependent radius is given by
\begin{equation} \label{eq:BP}
\tan \left( \frac{\Theta}{2} \right) = \frac{R(t)}{r}.
\end{equation}
This is valid from $r=0$ up to a distance $r \sim O(1/\ln R)$ when $R \ll 1$ (this is based on Eq.~(26) of Ref.~\cite{KomineasMelcherVenakides_PhysD2021}).
The kinetic energy is
\begin{equation} \label{eq:kinetic_small}
T = \frac{1}{2} \int \dot{\Theta}^2 \, 2\pi r\,dr \approx -4\pi \ln R\,\dot{R}^2
\end{equation}
where we have taken the limits in the integral \eqref{eq:kinetic_small} from $r=0$ to $r \sim 1/\ln R$ and we have only kept the dominant term for $R\to 0$.
The skyrmion profile decays exponentially for larger distances $r$ and we thus neglect this contribution to $T$.
The quantity $m=-8\pi \ln R$ can be considered as the mass of the breathing skyrmion and it is diverging for $R\to 0$.
This behavior is connected with the well-known divergence of the integrated magnetization for the BP skyrmion.
A nonzero kinetic energy implies
\begin{equation} \label{eq:Rdot}
    \dot{R} \sim \frac{1}{\sqrt{-\ln R}}, \qquad R\to 0.
\end{equation}
Based on Eq.~\eqref{eq:Rdot}, we anticipate that $\dot{R}\to 0$ as $R\to 0$ while at the same time the kinetic energy remains strictly positive due to the mass term.
We finally mention that the time it takes to achieve the annihilation via breathing is finite as can be found by integrating in time $\dot{R}$ in Eq.~\eqref{eq:Rdot}. 

\begin{figure}[t]
    \centering
    \includegraphics[width=\columnwidth]{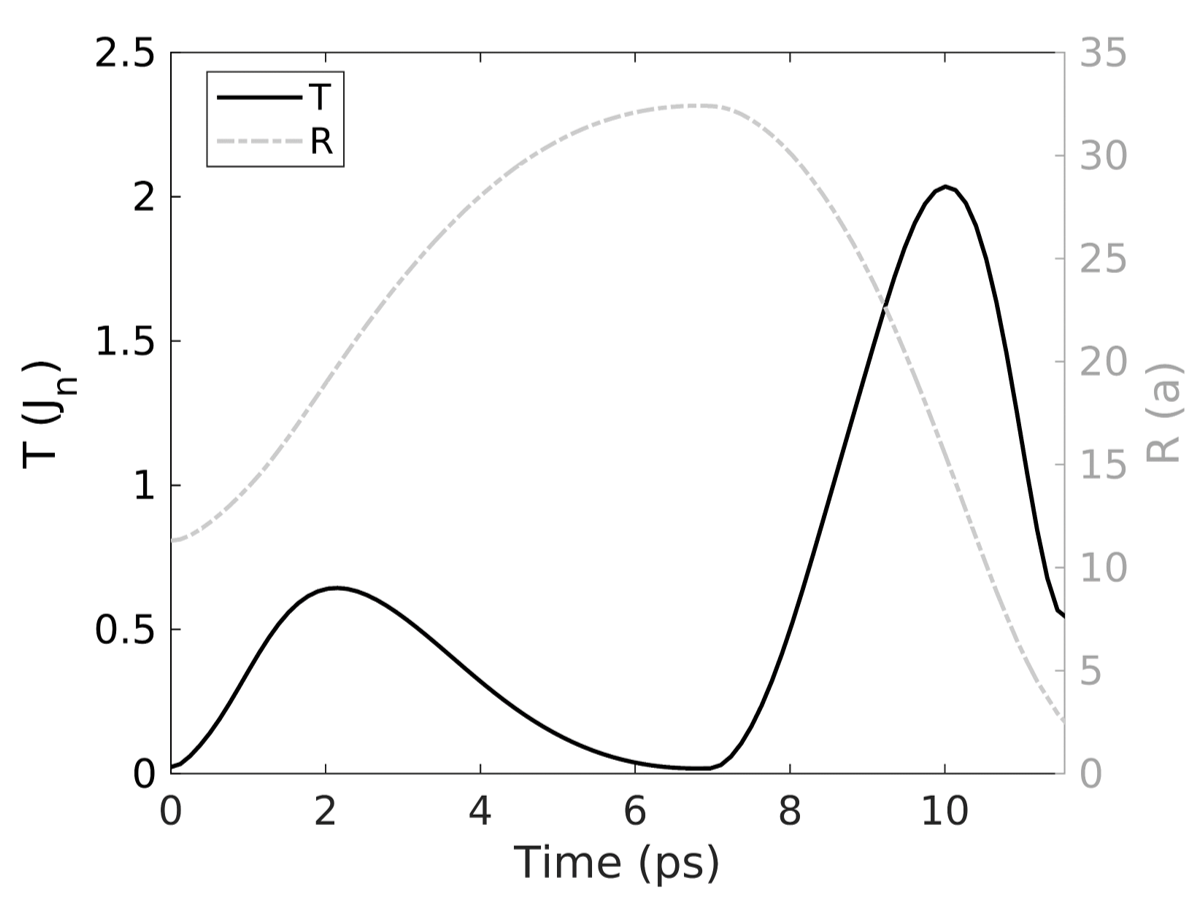}
    \caption{Kinetic energy ($T$) shown by a solid line for the annihilation event of Fig.~\ref{fig:annihilationDynamics}.
    The skyrmion radius ($R$) is also shown by a grey dashed line for comparison.
    The results are plotted up to the time point where $\Potential\approx 4\pi$, after which the continuum theory ceases to be valid.
    At this point, the skyrmion collapses while it can be clearly seen that $T > 0$, in agreement with the theoretical prediction.}
    \label{fig:annihilation_energy}
\end{figure}

Fig.~\ref{fig:annihilation_energy} shows the kinetic energy \eqref{eq:kinetic_magnetization_discrete} for the numerical simulation in Fig.~\ref{fig:annihilationDynamics} until the skyrmion is annihilated.
The skyrmion radius is shown in the same figure for comparison.
The kinetic energy starts from zero, reaches a maximum and returns to a very small value as the skyrmion radius approaches a maximum.
At that time, the pulse is switch off and the skyrmion radius starts decreasing rapidly until the skyrmion annihilates.
The kinetic energy has a nonzero value at the time of annihilation.
The numerical results confirm the predictions of the previous paragraphs that are based on the continuum model.

Methods usually employed for the creation or annihilation of complex topological textures involve driving them at material boundaries or forcing them to shrink down to the atomic size due to large fields.
By contrast, the annihilation of the skyrmion described in this section is obtained due to a mild perturbation.
Shrinking follows by the natural breathing dynamics and a singularity formation happens in a finite time interval.
It should also be noted that the nontrivial topology of the skyrmion is not an obstacle in the singularity formation and eventually in the annihilation process.

The proposed dynamics can be induced in various ways.
(i) One could temporarily increase the scaled parameter $\dm$ (by modifying the DM or anisotropy parameters), as presented in Fig.~\ref{fig:annihilationDynamics}.
(ii) One might also decrease the parameter, thus initiating breathing, but, in this case, a sufficiently low value should be maintained until the skyrmion annihilates.
(iii) Alternatively, an external magnetic field would act as easy-plane anisotropy in an antiferromagnet effectively reducing the easy-axis anisotropy parameter and thus increasing the dimensionless parameter $\dm$.
This case would require a separate study due to the more involved dynamics introduced by the external field \cite{BaryakhtarChetkin1994,KomineasPapanicolaou_NL1998}.
(iv) Modification of the kinetic energy of the skyrmion could also give further alternatives in order to initiate annihilation dynamics.
This could be achieved by inducing magnetization in the antiferromagnetic lattice.

\section{Nonlinear breathing mode}
\label{sec:nonlinear}

We proceed to a systematic study of the breathing dynamics in the linear and the nonlinear regime that will lead to quantitative predictions for the breathing and for the annihilation dynamics.
We consider, in this section, modes with helicity $\chi=0$ and a dynamic skyrmion profile
\begin{equation} \label{eq:skyrmionBrething}
    \Theta = \Theta(r,t),\quad \Phi=\phi.
\end{equation}
The kinetic energy in Eq.~\eqref{eq:kinetic_spherical} reduces to
\begin{equation} \label{eq:kinetic_chi0}
    T = \frac{1}{2} \int \dot{\Theta}^2 \, 2\pi r\,dr
\end{equation}
and the potential energy is
\begin{equation} \label{eq:potential_dm}
    \Potential_\dm(\Theta) = \Eex + \dm\, \Edmm + \Ean
\end{equation}
with $\Eex, \Edmm, \Ean$ defined in Eqs.~\eqref{eq:energyExchange_skyrmion}, \eqref{eq:energyDM_skyrmion}.

We aim to study breathing oscillations, that is, a periodic change of the skyrmion profile $\Theta(r,t)$.
We assume a material with parameter $\dm_0$.
In order to invoke breathing dynamics, we change the parameter to a value $\dm \ne \dm_0$ (for example, by applying a voltage) and assume that the skyrmion profile eventually relaxes to $\Theta_\dm$.
When the parameter is restored to the value $\dm_0$, breathing dynamics is initiated.

In order to make progress analytically, we make a simplifying assumption that is supported by numerical simulations.
As the skyrmion radius changes (oscillates) during breathing, we assume that the profile adiabatically adjusts to the static skyrmion profile $\Theta_\dm$ for the corresponding radius $R=R(\dm)$.
Under this assumption, the potential energy that the skyrmion experiences during the breathing motion is given by \eqref{eq:potential_dm} applied for $\dm=\dm_0$,
\begin{equation} \label{eq:potential_dm0}
\begin{split}
    \Potential_{\dm_0}(\Theta_\dm) & = \Eex + \dm_0 \Edmm + \Ean \\
    & = \Potential_\dm + (\dm_0-\dm) \Edmm
\end{split}
\end{equation}
where all terms are evaluated for $\Theta=\Theta_\dm$.
The energy components for each value of the parameter $\dm$ and the corresponding value of radius $R$ can be calculated numerically.

In Fig.~\ref{fig:breathingPotential}, we plot (by solid lines) the potential energy \eqref{eq:potential_dm0} for the eqilibrium profiles $\Theta_\dm$ as a function of their radius $R$.
Three lines are plotted for the cases $\dm_0=0.4, 0.47, 0.57$, which correspond to static skyrmion solutions with radius $R_0 = 0.71, 1.14, 2.39$ respectively.
As expected, every solid line has a minimum at the corresponding value of the radius. 

The main features of the potential wells in Fig.~\ref{fig:breathingPotential} can be anticipated.
We first consider the case of small radius, that is, $R\to 0$, or, equivalently, $\dm\to 0$ \cite{KomineasMelcherVenakides_NL2020,BernandMuratovSimon_PRB2020}.
We have $\Eex \to 4\pi,\, \Edm, \Ean \to 0$ and this gives $\Potential_\dm \to 4\pi$.
We also have $\Edmm \sim -8\pi R$, and thus Eq.~\eqref{eq:potential_dm0} gives
\begin{equation} \label{eq:potential_limit1}
    \Potential_{\dm_0} \to 4\pi,\quad\text{as}\quad R\to 0.
\end{equation} 
In the case of large radius, we have $\Potential_\dm \sim O(R^{-1})$ and $\Edm = -4\pi R$ as $R\to\infty$, or, equivalently, $\dm\to 2/\pi$ \cite{KomineasMelcherVenakides_PhysD2021}.
Thus, Eq.~\eqref{eq:potential_dm0} gives
\begin{equation} \label{eq:potential_limit2}
\Potential_{\dm_0} \sim  4\pi \left( 1 - \frac{\pi\dm_0}{2} \right) R,\quad\text{as}\quad R\to \infty
\end{equation}
where we have used $\dm = 2/\pi + O(R^{-2})$.
Eqs.~\eqref{eq:potential_limit1}, \eqref{eq:potential_limit2} confirm the features of the potentials shown in Fig.~\ref{fig:breathingPotential} that are crucial for describing breathing dynamics. 

\begin{figure}[t]
    \centering
    \includegraphics[width=\columnwidth]{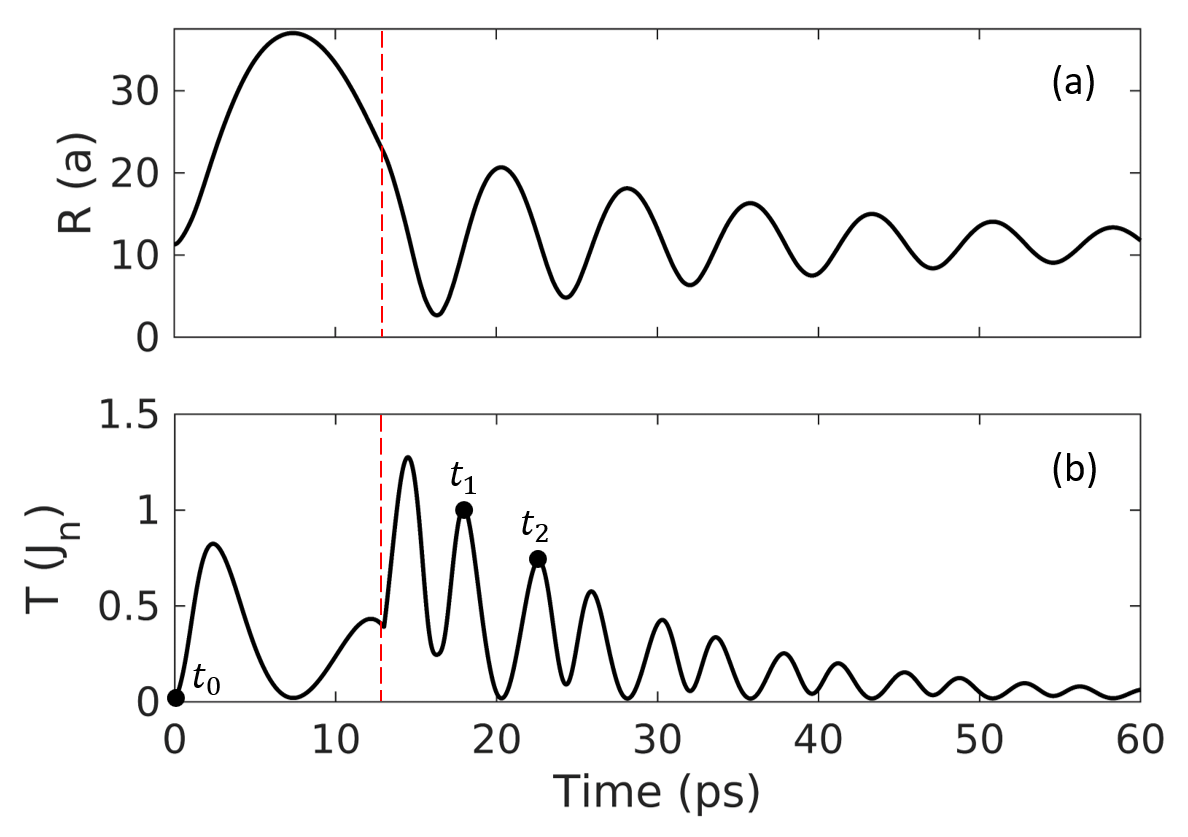}
    \caption{
    We apply a square pulse that modifies the DM parameter, as in Fig.~\ref{fig:annihilationDynamics}, but now with a duration of 13 ps.
    We start the simulation with $\DM=0.057$ ($\dm=0.57$) and we restore to $\DM=0.047$ ($\dm=0.47$) at the time marked by the dotted red line.
    The damping parameter is $\damp=0.001$.
    (a) Skyrmion radius vs time shows damped breathing oscillations.
    (b) Kinetic energy \eqref{eq:kinetic_magnetization_discrete}.
    The kinetic energy is close to zero at both turning points of the oscillation.}
    \label{fig:energyOscillation}
\end{figure}

The potential wells shown in Fig.~\ref{fig:breathingPotential} give rise to oscillating motion around the minimum $\Potential_0= \Potential_{\dm_0}(\Theta_{\dm_0})$.
We demonstrate this by performing a simulation similar to that in Sec.~\ref{sec:annihilation}, but we now retain the voltage pulse for a longer time, 13 ps.
We use a smaller damping parameter $\damp=0.001$ in order to demonstrate clearer the oscillating motion.
Fig.~\ref{fig:energyOscillation} shows the skyrmion radius and the kinetic energy during the motion.
An oscillating motion starts in the potential well $\Potential_{\dm_0=0.57}$.
The skyrmion radius has passed the maximum when the pulse is switched off.
Then, a new oscillating motion sets in in the potential well $\Potential_{\dm_0=0.47}$.
The oscillations eventually die out due to damping.
The kinetic energy $T$ is close to zero at both turning points of the oscillation, i.e., at the minimum and maximum radii.

\begin{figure*}[t]
    \centering
    \includegraphics[width=\textwidth]{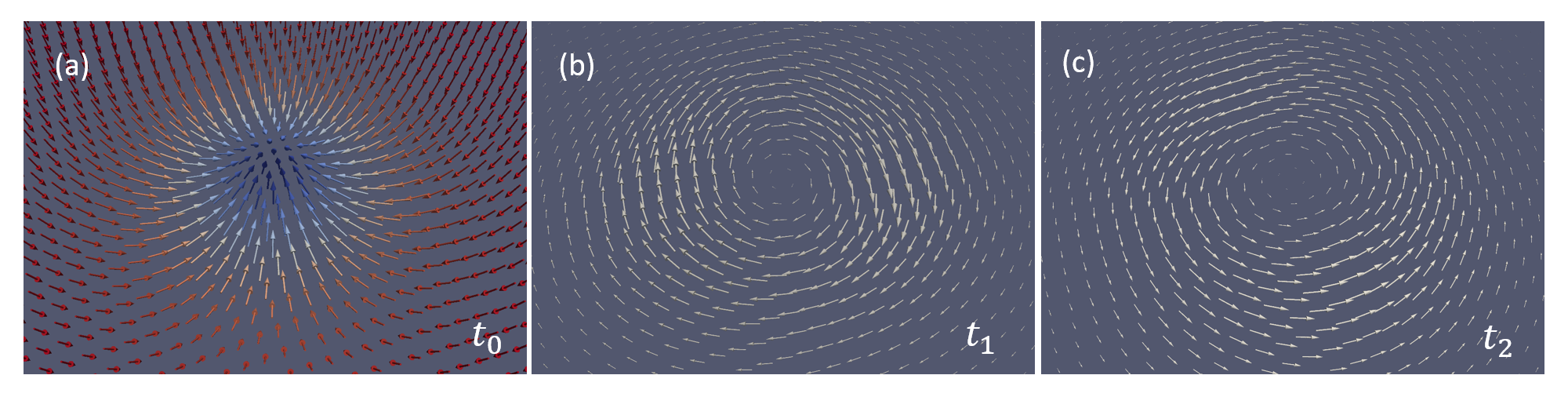}
    \caption{The N\'eel field $\nagn$ and the magnetization field $\magn$ of the skyrmion during breathing.
    (a) The N\'eel field for the initial skyrmion.
    (b) The magnetization field during skyrmion expansion, at time $t_1$ indicated in Fig.~\ref{fig:energyOscillation}.
    (c) The magnetization field during skyrmion shrinking, at time $t_2$.
    The magnetization points azimuthally in both cases as anticipated by Eq.~\eqref{eq:magnetization_chi0}.
}
    \label{fig:Magnetization}
\end{figure*}

Fig.~\ref{fig:Magnetization} shows snapshots of the N\'eel vector and the magnetization for the skyrmion.
The N\'eel vector $\nagn$ of the initial skyrmion is shown in entry (a).
The magnetization $\magn$ is shown during shrinking in entry (b) and expansion in entry (c).
It points azimuthally in the plane and gives opposite vectors during shrinking and expansion.
This is expected as the magnetization in Eq.~\eqref{eq:magn0} for a time dependent profile as in Eq.~\eqref{eq:skyrmionBrething} is
\begin{equation} \label{eq:magnetization_chi0}
\magn = \frac{\epsilon}{2\sqrt{2}}\,\dot{\Theta}\,\e_\phi.
\end{equation}

\begin{figure}[t]
    \centering
    \includegraphics[width=8cm]{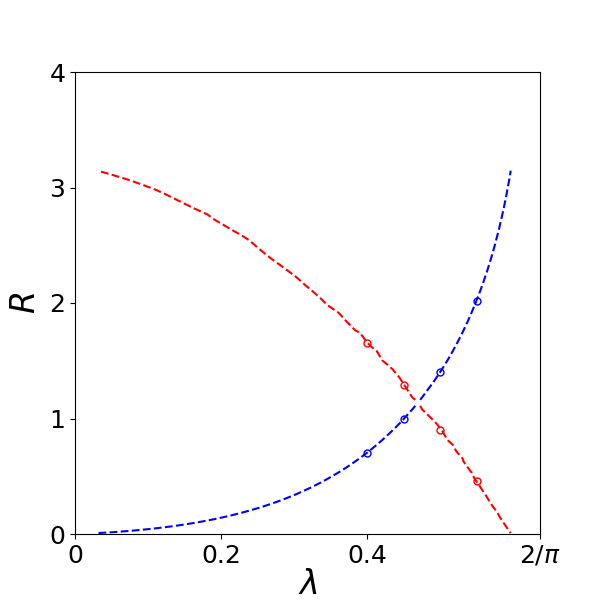}
    \caption{For $\dm_0=0.47$ ($\DM=0.047$), we show the maximum and minimum radii of skyrmion breathing oscillations, when the initial skyrmion profile is $\Theta_\dm$.
    Lines show theoretical results based on Eq.~\eqref{eq:potential_dm} and Fig.~\ref{fig:breathingPotential}.
    Circles show results of numerical simulations.
    The simulations are initiated with skyrmion profiles at the blue circles (they give the one extremum of the oscillation).
    The other extremum of the oscillation is shown by red circles.
    (The radius $R$ is in scaled units and the actual length is given, as in Eq.~\eqref{eq:physicalLength}, by $a R/\epsilon$, with $\epsilon=0.1$ for parameter values \eqref{eq:parameterValues}).
}
    \label{fig:nonlinear_Rmin-Rmax}
\end{figure}

We proceed to systematic simulations of large amplitude breathing oscillations.
We set $\DM=0.047$ ($\dm_0=0.47$) and we simulate the conservative equations, i.e., set $\damp=0$ in Eq.~\eqref{eq:Heisenberg}, for a range of initial profiles $\Theta_\dm$.
If the energy $\Energy$ is $\Potential_0 < \Energy < 4\pi$, we expect oscillating motion between the two values of the radius $R_{\rm min},\,R_{\rm max}$ for which the potential takes the value $\Potential_{\dm_0}=\Energy$.
It is $R_{\rm min} < R_0 < R_{\rm max}$ where $R_0$ is the skyrmion radius at the minimum of the potential.
We observe almost perfectly periodic motion that was verified for many periods.

Fig.~\ref{fig:nonlinear_Rmin-Rmax} shows numerical results for the amplitude of oscillations, at $\dm_0=0.47$, for simulations initiated with profiles $\Theta_\dm$ for a range of values of the parameter $\dm$.
The radius of the initial profiles are plotted by blue circles and they represent the one extremum of the oscillating motion.
Red circles give the radius of the skyrmion at the other extremum of the oscillation.
The blue line shows the skyrmion radius corresponding to the static profile $\Theta_\dm$ and the red line the radius for the partner configuration with the same potential energy $\Potential$.
If we choose an initial profile $\Theta_\dm$ with $\dm > 0.6$, we have $V > 4\pi$ and the skyrmion annihilates during the shrinking phase, as discussed in Sec.~\ref{sec:annihilation}.
Thus, the numerical results are in very good agreement with the assumption that the oscillating motion is described by the potentials in Fig.~\ref{fig:breathingPotential}.

The period of oscillation for a small amplitude is found to be approximately $T=20$ in dimensionless units (or $T=7.5\,{\rm ps}$ when we use the time unit in Eq.~\eqref{eq:physicalTime}) that corresponds to an angular frequency $\omega_b=0.3$.
This is in agreement with the analytical result \eqref{eq:frequency_large} given in the next section.
The frequency is decreasing for large amplitude oscillations.
This is anticipated since the potential in Fig.~\ref{fig:breathingPotential} is slower than parabolic.


We return to the issue of skyrmion annihilation and we can now expand upon the results of Sec.~\ref{sec:annihilation}.
Based on the potential wells shown in Fig.~\ref{fig:breathingPotential}, one can start breathing dynamics either pushing towards a radius smaller than that at the potential minimum or to a larger radius (as we have shown in Fig.~\ref{fig:annihilationDynamics}).
Furthermore, one can imagine a combination of the two possibilities.
That is, one may modulate $\dm$ periodically around the value $\dm_0$, thus pushing the skyrmion radius to values smaller and larger than $R_0$ periodically.
Using the appropriate frequency for the $\dm$ modulation, this is expected to lead to resonance, i.e., large amplitude oscillations for the skyrmion radius, and to its eventual annihilation.
A small amplitude modulation of $\dm$ will be sufficient for the resonance phenomenon.

\section{Small breathing oscillations}
\label{sec:smallOscillations}

The breathing motion with a small amplitude will be studied analytically, based on the potential \eqref{eq:potential_dm0} plotted in Fig.~\ref{fig:breathingPotential}.
We will study separately the case of a skyrmion of large radius and a skyrmion of small radius.

\subsection{Skyrmions of large radius}
\label{sec:largeRadius}

Let us consider a large value of $\dm_0$ so that the corresponding static skyrmion has a large radius.
The skyrmion radius $R(t)$ will vary with time during the breathing motion.
The skyrmion profile is approximated as a domain wall centered at the position of the radius, 
\begin{equation}
\Theta = 2 \arctan\left( e^{-z} \right),\qquad z\equiv r-R(t)
\end{equation}
and we assume that $\chi=0$ during the motion.
The kinetic energy \eqref{eq:kinetic_spherical} gives
\[
T = \frac{\dot{R}^2}{2} \int_0^\infty \sech^2(r-R) \, (2\pi r\,dr).
\]
For $R\gg 1$, we may extend the lower limit to $-\infty$ (with an exponentially small error) and obtain
\begin{equation} \label{eq:kinetic_large}
T \approx \frac{\dot{R}^2}{2} \int_{-\infty}^\infty \sech^2 z \, (2\pi R\,dz) = 2\pi R \dot{R}^2.
\end{equation}

For calculating the potential energy \eqref{eq:potential_dm0} we will use the results \cite{KomineasMelcherVenakides_PhysD2021}
\begin{equation} \label{eq:energyAsymptotic_large}
\begin{split}
V_\dm & = \frac{4\pi^2 \dm_2}{R} + O\left( R^{-3} \right),\quad \dm_2 = 0.3057 \\
\Edm & = -4\pi R + O\left( R^{-1} \right)
\end{split}
\end{equation}
where $R$ is the radius of the static skyrmion for DM parameter $\dm$, approximated by \cite{KomineasMelcherVenakides_PhysD2021}
\begin{equation} \label{eq:dm-R_large}
\dm = \frac{2}{\pi} - \frac{\dm_2}{R^2} + O\left( R^{-4} \right).
\end{equation}
Our objective is to evaluate the potential \eqref{eq:potential_dm0} around the radius $R_0$ that corresponds to the parameter $\dm_0$.
We set
\begin{equation} \label{eq:delta}
R = R_0(1+\delta)
\end{equation}
and we use Eq.~\eqref{eq:dm-R_large} to obtain
\begin{equation} \label{eq:dm-dm0}
\frac{\dm_0}{\dm} - 1 \approx -\frac{\pi \dm_2}{R_0^2} \left( 1 - \frac{3}{2}\delta \right) \delta,\quad \delta \ll 1.
\end{equation}
Inserting Eq.~\eqref{eq:dm-dm0} and \eqref{eq:delta} in Eq.~\eqref{eq:energyAsymptotic_large}, we have
\begin{align*}
& V_\dm \approx \frac{4\pi^2 \dm_2}{R_0} ( 1 - \delta + \delta^2) \\
& \frac{\dm_0-\dm}{\dm}\, \Edm
\approx \frac{4\pi^2 \dm_2}{R_0} \left( \delta - \frac{1}{2}\delta^2 \right).
\end{align*}
Substituting the two last equations in the potential energy \eqref{eq:potential_dm0}, we obtain the parabolic form
\begin{equation} \label{eq:potential_dm0_large}
V_{\dm_0}(\Theta_\dm) = \frac{4\pi^2 \dm_2}{R_0} \left( 1 + \frac{1}{2}\delta^2 \right).
\end{equation}

Eqs.~\eqref{eq:kinetic_large}, \eqref{eq:potential_dm0_large} give the Lagrangian
\begin{equation} \label{eq:Lagrangian_large}
L = 2\pi R_0^3 \dot{\delta}^2 - \frac{2\pi^2 \dm_2}{R_0} \delta^2
\end{equation}
which implies harmonic oscillations with angular frequency
\begin{equation} \label{eq:frequency_large}
\omega_b = \frac{\sqrt{\pi \dm_2}}{R_0^2}
\approx \sqrt{\frac{\pi}{\dm_2}} \left( \frac{2}{\pi} - \dm_0 \right).
\end{equation}

\begin{figure}[t]
    \centering
    \includegraphics[width=7cm]{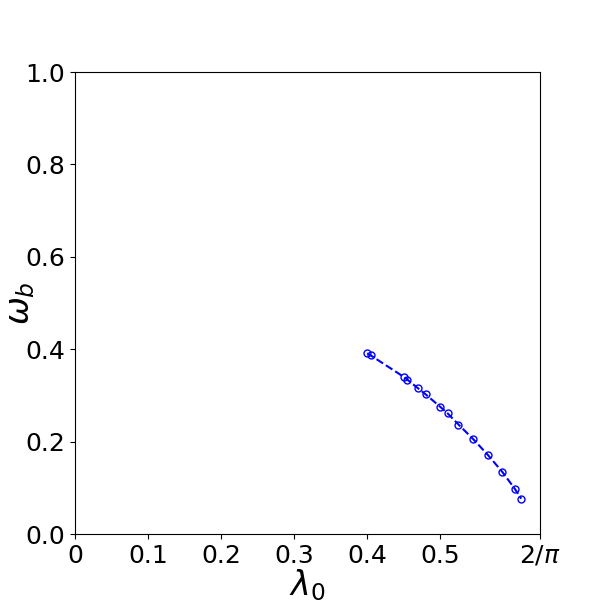}
    \caption{The angular frequency for small amplitude breathing oscillations as a function of the parameter $\dm_0$.
    The angular frequency goes to zero as $\dm_0$ approaches $2/\pi$.
    The rate of convergence is consistent with Eq.~\eqref{eq:frequency_large} within numerical accuracy.
    Physical units for frequency $\omega_b/(2\pi)$ are restored by multiplying by $2.68\,{\rm THz}$. 
    }
    \label{fig:linear_frequency}
\end{figure}

Fig.~\ref{fig:linear_frequency} shows the results of numerical simulations for breathing oscillations of small amplitude.
For $\dm_0$ close to the value $2/\pi$, the numerical results confirm Eq.~\eqref{eq:frequency_large}.
The dependence of the breathing frequency on $R_0$ has been obtained in Ref.~\cite{2019_PRB_KravchukGomonaySheka}, but the numerical factor in Eq.~\eqref{eq:frequency_large} introduces a correction to that result.
This modification originates in the corrected dependence of the radius on the DM parameter given in Eq.~\eqref{eq:dm-R_large} compared to an earlier result obtained in Ref.~\cite{RohartThiaville_PRB2013}.

\subsection{Skyrmions of small radius}
\label{sec:smallRadius}

The profile of a skyrmion of small radius is approximated by a Belavin-Polyakov (BP) solution \cite{KomineasMelcherVenakides_NL2020,BernandMuratovSimon_PRB2020}.
We thus consider a BP skyrmion with a time dependent radius, as in Eq.~\eqref{eq:BP}.
The kinetic energy is given in Eq.~\eqref{eq:kinetic_small} or, using Eq.~\eqref{eq:delta},
\begin{equation} \label{eq:kinetic_small2}
T \approx -4\pi R_0^2 \ln R_0\, \dot{\delta}^2.
\end{equation}

Regarding the potential energy, we use the formula \eqref{eq:potential_dm0_small_corrected} derived in Appendix~\ref{sec:approximation_small}.
That is an improvement of the asymptotic results of Ref.~\cite{KomineasMelcherVenakides_NL2020}.
The parameter $\dm$ and the skyrmion radius $R$ for small radius are related by \cite{KomineasMelcherVenakides_NL2020,BernandMuratovSimon_PRB2020}
\begin{equation} \label{eq:dm-R_small}
    \dm = -R\ln R.
\end{equation}
Using the latter, the potential \eqref{eq:potential_dm0_small_corrected} is written as
\begin{equation} \label{eq:potential_dm0_small-R-R0}
V_{\dm_0}(R) = 4\pi  \left[ 1 - R (R\ln R - 2R_0\ln R_0) + \frac{1}{2} R^2 \right]
\end{equation}
that is valid for small $R$ and $R_0$.
Eq.~\eqref{eq:potential_dm0_small-R-R0} has a minimum at $R=R_0$ as desired (note that the minimum is obtained thanks to Eq.~\eqref{eq:potential_dm0_small_corrected} while it would not have been possible to obtained by using previously available formulas).
Inserting \eqref{eq:delta} in \eqref{eq:potential_dm0_small-R-R0}, we find the quadratic approximation
\begin{equation} \label{eq:potentialQuadratic_small}
V_{\dm_0} \approx 4\pi \left(1 + R_0^2 \ln R_0 - R_0^2 \ln R_0\, \delta^2 \right).
\end{equation}

Eqs.~\eqref{eq:kinetic_small2} and \eqref{eq:potentialQuadratic_small} give the Lagrangian
\begin{equation}
    L = -4\pi R_0^2 \ln R_0\, (\dot{\delta}^2 - \delta^2).
\end{equation}
We thus find that small breathing oscillations for skyrmions of small radius have an angular frequency
\begin{equation} \label{eq:frequency_small}
    \omega_b = 1.
\end{equation}
Note that this is equal to the frequency \eqref{eq:frequency_helicity} for helicity oscillations discussed in the next Section.

The numerical results shown in Fig.~\ref{fig:linear_frequency} indicate that $\omega_b$ takes a value of $O(1)$ for skyrmions of small radius and it is thus consistent with the result in Eq.~\eqref{eq:frequency_small}.
We cannot simulate the dynamics of skyrmions for very small $\dm$ due to their very small size and we thus do not present a precise numerical result for $\omega_b$ as $\dm_0\to 0$.

\section{Helicity oscillations}
\label{sec:helicity}

For completeness, we consider an oscillation mode where the radial skyrmion profile remains unchanged while the helicity depends on time
\begin{equation} \label{eq:helicityOscillation}
    \Theta = \Theta(r),\quad \Phi = \phi + \chi(t).
\end{equation}
However, this assumption is not consistent with the equations of motion derived from the energy functional.
Indeed, simulations show that Eq.~\eqref{eq:helicityOscillation} does not give a good approximation for the dynamical profile.
When we start from a static skyrmion profile $\Theta_\dm$ and choose uniform helicity $\chi \ne 0$, we obtain a complicated motion that seems to combine breathing and helicity oscillations.
Nevertheless, Eq.~\eqref{eq:helicityOscillation} will prove its usefulness as it will lead to an approximation for the equation of motion for $\chi$ and to the frequency of the observed oscillations.
We therefore argue that this assumption is useful and we proceed to use it in the following calculations.

The kinetic energy \eqref{eq:kinetic_spherical} reduces to the expression
\begin{equation}
T = \frac{1}{2} \dot{\chi}^2 \int \sin^2\Theta\, (2\pi r\,dr) = \dot{\chi}^2 \Ean
\end{equation}
and the potential energy is
\begin{equation}
    \Potential_\dm = \Eex + \dm\cos\chi\, \Edmm + \Ean
\end{equation}
where $\Eex, \Edmm, \Ean$ depend on $\Theta$ but not on $\chi$, as seen in Eqs.~\eqref{eq:energyExchange_skyrmion}, \eqref{eq:energyDM_skyrmion}.
Omitting terms independent of $\chi$, the Lagrangian is
\begin{equation} \label{eq:Lagrangian_helicity0}
    L = T - \Potential = \dot{\chi}^2 \Ean - \dm \cos\chi\, \Edmm.
\end{equation}
By a standard scaling argument for the minimizer $\Theta_\dm$, a virial relation is obtained \cite{BogdanovHubert_PSS1994},
\begin{equation} \label{eq:virial}
    2\Ean + \Edm = 0 \Rightarrow \dm\Edmm = - 2\Ean.
\end{equation}
The latter is used in Eq.~\eqref{eq:Lagrangian_helicity0} to give
\begin{equation} \label{eq:Lagrangian_helicity}
    L = 2\Ean \left( \frac{1}{2} \dot{\chi}^2 + \cos\chi \right)
\end{equation}
Lagrangian \eqref{eq:Lagrangian_helicity} describes a pendulum.
For small $\chi \ll 1$, it gives harmonic oscillations with angular frequency
\begin{equation} \label{eq:frequency_helicity}
    \omega_h = 1
\end{equation}
that is a period $T=2\pi$ (or $T=2.3\,{\rm ps}$ when we use the time unit in Eq.~\eqref{eq:physicalTime}).
Large amplitude oscillations will have a smaller frequency as in the case of a pendulum.
The value in Eq.~\eqref{eq:frequency_helicity} agrees with the results of numerical simulations.

Note that oscillations of helicity, for a small amplitude, are found to have the same frequency \eqref{eq:frequency_helicity} as small breathing oscillations for skyrmions of small radius \eqref{eq:frequency_small}.


\section{Concluding remarks}
\label{sec:conclusions}

We have studied breathing oscillations of skyrmions in chiral antiferromagnets using analytical arguments and calculations, within a continuum model, that are valid in the nonlinear regime.
The predictions are confirmed and the results are extended by systematic numerical simulations within the original discrete spin model.
A significant result is the prediction, confirmed by simulations, that the forced expansion of the skyrmion radius invokes breathing oscillations that can lead to the skyrmion annihilation.
This counter-intuitive process offers an alternative to the typically employed methods of forced skyrmion suppression.
It can prove to be a more convenient method as it only requires mild forcing.
This is sufficient because the annihilation is actually brought about, not by the forcing itself, but by the invoked internal dynamics.
Furthermore, the phenomenon is interesting also from a theoretical perspective because it involves the creation of a singularity in finite time.
Finally, we give an analytical calculation based on an energetic method for the frequency of small amplitude oscillations.

In the development of the theoretical arguments, we have given details about the kinetic energy of the continuum model for an antiferromagnet.
This is actually an emergent kinetic energy that originates in the exchange energy of the discrete model.
We derive the surprising result that the kinetic energy can be tuned to a nonzero value at the point of the singularity formation and skyrmion annihilation.

Given the counter-intuitive dynamical behaviour of the breathing skyrmion, it will be interesting to consider further dynamical phenomena of this system.
For example, it is interesting to investigate the skyrmion domain wall velocity during breathing.
If this could reach the maximum velocity allowed for a traveling wall, that is $v_{\rm max} = \sqrt{1-(\pi\dm/2)^2}$ \cite{2021_PRB_TomaselloKomineas}, then an instability of the N\'eel state will occur that may lead to the spontaneous formation of a spiral.

\section*{Acknowledgements}
This work was supported by the project “ThunderSKY” funded by the Hellenic Foundation for Research and Innovation and the General Secretariat for Research and Innovation, under Grant No. 871.

\appendix

\section{Approximation of the potential for small radius}
\label{sec:approximation_small}

The available asymptotic formulae for the energy at small $\dm$ are \cite{KomineasMelcherVenakides_NL2020}
\begin{equation} \label{eq:energyAsymptotic_small}
\begin{split}
& V_\dm(\Theta_\dm) = 4\pi \left( 1 + \frac{\dm^2}{\ln \dm} \right)  + o\left( \frac{\dm^2}{\ln \dm}  \right),\\
& \Edm = 8\pi\,\frac{\dm^2}{\ln \dm} + o\left( \frac{\dm^2}{\ln \dm}  \right).
\end{split}
\end{equation}
Substituting in the potential energy \eqref{eq:potential_dm0}, we obtain
\begin{equation} \label{eq:potential_dm0_small-R}
V_{\dm_0} = 4\pi  \left[ 1 + (2\dm_0 - \dm) \frac{\dm}{\ln\dm} \right].
\end{equation}
The derivative of this potential is
\begin{equation} \label{eq:Vp_dm0-1}
V_{\dm_0}' = 4\pi \left[ 2\frac{\dm_0 - \dm}{\ln\dm} - \frac{2\dm_0-\dm}{(\ln\dm)^2} \right].
\end{equation}
We have $V_{\dm_0}'(\dm_0) \ne 0$, due to the term $1/(\ln\dm)^2$, and thus no minimum is implied at $\dm=\dm_0$.
The only way to obtain a formula that gives $V_{\dm_0}'(\dm_0) = 0$ up to terms $1/(\ln\dm)^2$ is to write
\begin{equation} \label{eq:energy_corrected}
\Eex = 4\pi\left[ 1 + \frac{1}{2} \frac{\dm^2}{(\ln \dm)^2} \right],\qquad
\Edm = 8\pi\, \frac{\dm^2}{\ln \dm},
\end{equation}
that gives
\begin{equation} \label{eq:potential_dm0_small_corrected}
V_{\dm_0}(\dm) = 4\pi \left[ 1 + (2\dm_0 - \dm) \frac{\dm}{\ln\dm} + \frac{1}{2} \frac{\dm^2}{(\ln \dm)^2} \right].
\end{equation}
We have
\begin{equation}
V_{\dm_0}' = 4\pi  \left[ 2\frac{\dm_0 - \dm}{\ln\dm} + 2 \frac{\dm_0-\dm}{(\ln\dm)^2} \right] + O\left( \frac{\dm}{(\ln \dm)^3} \right)
\end{equation}
and thus $V_{\dm_0}$ has the desired behaviour at $\dm=\dm_0$.

Formula \eqref{eq:energy_corrected} for $\Eex$ is an improvement over previous results \cite{KomineasMelcherVenakides_PhysD2021} and it is necessary in order to find the oscillation frequency for breathing oscillations in Sec.~\ref{sec:smallRadius}.

\bigskip


\end{document}